\documentclass[a4paper]{article}
\usepackage{amsmath,amssymb,theorem,fullpage,mathtools}
\usepackage[shortcuts]{extdash}
\usepackage{hyperref}
\usepackage[nosort]{cite}
\usepackage[shortcuts]{extdash}

\let\frac\undefined

\allowdisplaybreaks

\numberwithin{equation}{section}

\def\Maketitle{{\def\newpage{}\maketitle}}

\def\eq#1{\begin{equation}#1\end{equation}}
\long\def\subeq#1{\begin{subequations}#1\end{subequations}}

\def\Align#1{\begin{align}#1\end{align}}

\def\Aligned#1{\begin{aligned}#1\end{aligned}}
\def\Gather#1{\begin{gather}#1\end{gather}}
\def\Gathered#1{\begin{gathered}#1\end{gathered}}
\def\Multline#1{\begin{multline}#1\end{multline}}

\def\Cases#1{\begin{cases}#1\end{cases}}
\def\?{\notag}
\def\d{\partial}
\def\bd{\bar\partial}

\def\Res{\mathop{\rm Res}}

\def\cA{{\cal A}}

\def\cD{{\cal D}}
\def\cF{{\cal F}}
\def\bcF{{\cal\bar F}}

\def\cO{{\cal O}}

\def\sh{\mathop{\rm sh}\nolimits}
\def\ch{\mathop{\rm ch}\nolimits}
\def\th{\mathop{\rm th}\nolimits}

\def\lcolon{\mathopen{\,:\,}}
\def\rcolon{\mathclose{\,:\,}}

\def\Z{{\mathbb Z}}

\def\e{{\rm e}}
\def\i{{\rm i}}

\def\tB{{\tilde B}}

\def\tF{{\tilde F}}

\def\tS{{\tilde S}}
\def\tq{{\tilde q}}
\def\tQ{{\tilde Q}}
\def\tsigma{{\tilde\sigma}}
\def\tSigma{{\tilde\Sigma}}
\def\tW{{\tilde W}}
\def\tdelta{{\tilde\delta}}

\def\bh{{\bar h}}

\makeatletter

\def\section{\@startsection{section}{1}{\z@}%
                                   {-3.5ex \@plus -1ex \@minus -.2ex}%
                                   {2.3ex \@plus.2ex}%
                                   {\normalfont\normalsize\bfseries}}
\def\subsection{\@startsection{subsection}{2}{\z@}%
                                     {-3.25ex\@plus -1ex \@minus -.2ex}%
                                     {1.5ex \@plus .2ex}%
                                     {\normalfont\normalsize\bfseries\itshape}}
\def\@seccntformat#1{\csname the#1\endcsname.~~}
\long\def\@makecaption#1#2{%
  \vskip\abovecaptionskip
  \sbox\@tempboxa{\small#1. #2}%
  \ifdim \wd\@tempboxa >0.9\hsize
  {\leftskip=0.05\hsize\rightskip=0.05\hsize\relax\small
    #1. #2\par}
  \else
    \global \@minipagefalse
    \hb@xt@\hsize{\hfil\box\@tempboxa\hfil}%
  \fi
  \vskip\belowcaptionskip}
\def\Appendix{\appendix
  \def\@seccntformat##1{Appendix~\csname the##1\endcsname.~~}}

\let\over\@@over
\let\atop\@@atop
\let\above\@@above
\let\overwithdelims\@@overwithdelims
\let\atopwithdelims\@@atopwithdelims
\let\abovewithdelims\@@abovewithdelims

\makeatother

\begin{document}

\title{Form factors of descendant operators:\\ Resonance identities in the sinh-Gordon model}
\author{Michael Lashkevich${}^{1,2,3}$ and Yaroslav Pugai${}^{1,2}$,\\[\medskipamount]
\parbox[t]{0.9\textwidth}{\normalsize\it\raggedright
${}^1$~Landau Institute for Theoretical Physics, 142432 Chernogolovka, Russia\medspace%
\footnote{Mailing address.}
\\
${}^2$~Moscow Institute of Physics and Technology, 141707 Dolgoprudny, Russia\\
${}^3$~Kharkevich Institute for Information Transmission Problems, 19 Bolshoy Karetny per., 127994 Moscow, Russia}
}
\date{}

\Maketitle

\begin{abstract}
We study the space of local operators in the sinh\-/Gordon model in the framework of the bootstrap form factor approach. Our final goal is to identify the operators obtained by solving bootstrap equations with those defined in terms of the Lagrangian field. Here we try to identify operators at some very particular points, where the phenomenon of operator resonance takes place. The operator resonance phenomenon being perturbative, nevertheless, results in exact identities between some local operators. By applying an algebraic approach developed earlier for form factors we derive an infinite set of identities between particular descendant and exponential operators in the sinh\-/Gordon theory, which generalize the quantum equation of motion. We identify the corresponding descendant operators by comparing them with the result of perturbation theory.
\end{abstract}

\section{Introduction}

In this paper we continue developing the algebraic method for studying the structure of the space of local operators in two\-/dimensional integrable massive field theories, started in~\cite{Feigin:2008hs} and continued in~\cite{Alekseev:2009ik,Lashkevich:2013mca,Lashkevich:2013yja,Alekseev:2011my,Alekseev:2012jd}.

A description of the space of local operators is a necessary step for finding exact solutions to integrable models of quantum field theory. In a variety of models of conformal field theory the space of local operators (as well as the space of states) is completely described in terms of the representation theory of chiral algebras, and in the simplest case of the Virasoro algebra\cite{Belavin:1984vu}. This description makes it possible to find exact expressions for correlation functions. One of the most useful methods of calculating correlation functions is the free field representation\cite{Dotsenko:1984nm,Dotsenko:1984ad}, where the conformal blocks are represented in terms of an auxiliary free field. In the case of massive integrable field theory there is no algebra acting in the physical space similar to the chiral algebras in the conformal field theory. Exact expressions for correlation functions are only known in some very particular cases that are equivalent to free field theories, such as the Ising model\cite{Wu:1975mw}. Nevertheless, integrability imposes very strong restrictions on the $S$ matrix and form factors of local operators. Owing to this fact the $S$ matrices of many integrable theories including the sinh\-/Gordon theory are known exactly\cite{Vergeles:1976ra,Zamolodchikov:1978xm}. For any given model, the form factors of local operators are defined as matrix elements in the basis of eigenstates of the Hamiltonian. In integrable cases they satisfy a set of linear difference equations called form factor axioms\cite{Karowski:1978vz,Smirnov:1984sx,Smirnov:1992vz}. For a given integrable field theory every solution to these equations corresponds to a unique local operator. The form factors can also be expressed in terms of a free field theory in an auxiliary space of the spectral parameter\cite{Lukyanov:1993pn,Lukyanov:1993hc}. From the algebraic point of view this construction deforms the free field representation of the conformal field theory\cite{Lukyanov:1995gs}.

The infrared (large distance) asymptotics of the correlation functions can be found in terms of the form factors. The ultraviolet (small distance) asymptotics can be found independently of the integrability by means of conformal perturbation theory\cite{Zamolodchikov:1990bk}. In principle, both expansions converge fast enough to interpolate with an acceptable numeric precision. The main obstacle is that there is no natural way to identify particular fields in the Lagrangian theory as solution to the form factor equations. In a number of cases such an identification was found. In particular, in the sinh\-/Gordon theory this identification was established for the so called exponential fields (see below)\cite{Koubek:1993ke,Lukyanov:1997bp}. For general local operators no answer is known. Some proposals for the representation of general operators were proposed\cite{Koubek:1993ke,Babujian:2002fi}. A free field construction developing the Ansatz of~\cite{Babujian:2002fi} was proposed in~\cite{Feigin:2008hs}. In~\cite{Lashkevich:2013yja} it was shown that it is, in fact, based on a special limit of the deformed Virasoro algebra\cite{Shiraishi:1995rp}. An important ingredient of the free field description of the deformed Virasoro algebra are the so called screening operators. In fact, the deformed Virasoro algebra can be defined as the current algebra that commutes with the screening operators~\cite{Lukyanov:1994re,Lukyanov:1996qs}. It was shown in~\cite{Lashkevich:2013mca,Lashkevich:2013yja} that the existence of screening operators results in the emergence of identities between form factors of different local operators. Here we will discuss a particular class of these identities, which correspond to the so called resonance identities in conformal perturbation theory.

The operator resonance phenomenon was first described in~\cite{Zamolodchikov:1990bk}. It is related to ultraviolet divergences in conformal perturbation theory. If a local operator $\cO_0(x)$ of conformal dimension $\Delta$ is mixed in the $k$th order of perturbation theory with another operator $\cO_1(x)$ of conformal dimension $\Delta_1$, such that $\Delta_0>\Delta_1+k(1-\Delta_{\text{pert}})$, where $\Delta_{\text{pert}}$ is the dimension of the perturbation field, the perturbative integrals in correlation functions of the operator $\cO_0(x)$ diverge at small scales and a renormalization of this operator is necessary. The renormalized operator has the form $\cO^{\text{ren}}_0=\cO_0+C\cO_1$, where $C$ is a divergent $c$-numeric coefficient. In the marginal case $\Delta_0=\Delta_1+k(1-\Delta_{\text{pert}})$ the coefficient $C$ diverges as a logarithm of the ultraviolet cutoff $\log\Lambda$. The cutoff $\Lambda$ under the logarithm must be nondimensionalized by some arbitrarily chosen mass scale~$\lambda$. This results in an ambiguity in the definition of the renormalized operator~$\cO^{\text{ren}}$, which depends on~$\lambda$. If the operator $\cO_0$ is a member of a continuous family of operators $\cO_\Delta$, the correlation functions $\langle\cO_\Delta(x)\cdots\rangle$ all possess a pole at the point $\Delta=\Delta_0$. The residue of this pole is proportional to $\langle\cO_1(x)\cdots\rangle$ so that we may identify, up to a constant factor, $\Res_{\Delta=\Delta_0}\cO_\Delta(x)$ with the operator $\cO_1(x)$. Identities of such a form are called {\it resonance identities}. The physical renormalized operator $\cO_0^{\text{ren}}$ can be defined as the finite part of the $\Delta-\Delta_0$ expansion of $\cO_\Delta$. This finite part is again defined up to addition of $\cO_1(x)\log\lambda$. In the present article we will only be interested in the uniquely defined singular part.

The importance of the resonance phenomenon for the theory of form factors was first noticed in~\cite{Fateev:2009kp}. It was found that in the perturbed $Z_N$-parafermion models, the form factors of some operators, obtained by the use of standard general formulas, exactly vanish. This was shown to be compensated by infinite normalization factors, so that the vanishing form factors were interpreted as a result of resonance identities, and small deviations from the resonance points were necessary to obtain physical operators. On the other hand, resonance identities turn out to be very useful for the bootstrap theory of form factors. Indeed, resonance identities are perturbatively exact, so that they can be compared to exact identities for form factors obtained in the framework of the bootstrap theory.

In the present paper we find a wide class of resonance identities for form factors in the sinh-Gordon model, which generalize the identities obtained in~\cite{Lashkevich:2013mca,Lashkevich:2013yja}. To do it we study further properties of the algebra of screening operators, which have not been taken into account before. By comparing these results to the perturbative ones\cite{Lashkevich:2011ne} we will identify some particular operators defined in terms of form factors with those defined in terms of the Lagrangian field.

The sinh\-/Gordon model is a model of a real scalar boson field $\varphi(x)$ with action
\eq{
S[\varphi]=\int d^2x\,\left({(\d_\nu\varphi)^2\over16\pi}-\mu\e^{b\varphi}-\mu\e^{-b\varphi}\right).
\label{shG-action}
}
Here $b$ is a dimensionless parameter, which determines the dynamics of the system, while $\mu$ is a dimensional parameter with dimension $(\text{mass})^{2+2b^2}$. The space of local operators of the model consists of exponential operators $\e^{\alpha\varphi(x)}$ and their Fock descendants:
\eq{
\d^{k_1}\varphi\,\cdots\d^{k_s}\varphi\,\bd^{l_1}\varphi\,\cdots\,\bd^{l_t}\varphi\,\e^{\alpha\varphi},
\label{descendants-def}
}
where $\d={1\over2}(\d_x-\d_t)$, $\bd={1\over2}(\d_x+\d_t)$. The operators (\ref{descendants-def}) form a Fock space generated from the exponential operator by the Laurent components of the field $\varphi(x)$. The pair of nonnegative integers $(L,\bar L)=\left(\sum k_i,\sum l_i\right)$ is called the {\it level} of the descendant operator. Since in this paper we mostly consider spinless operators, for which $\bar L=L$, we refer to the single integer $L$ as the level of a spinless descendent operator.

This picture is based on the massless free field theory at the point $\mu=0$. The perturbation essentially modifies the structure of the space of local operators. Due to the specific form of the perturbation not all local operators are independent. First, the fields with different values of $\alpha$ are related by the so called reflection relations~\cite{Zamolodchikov:1995aa,Fateev:1997nn}. Second, note that the basis (\ref{descendants-def}) does not contain the mixed derivatives $\d^k\bd^l\varphi$ ($k,l>0$). In the classical theory they are excluded by means of the equation of motion
\eq{
\d\bd\varphi=2\pi\mu b\left(\e^{b\varphi}-\e^{-b\varphi}\right),
\label{shG-eqmotion}
}
so that instead of mixed derivatives we obtain fields with shifted parameter~$\alpha$. In the quantum theory we do not have this possibility. Nevertheless, at generic points of $\alpha$, conformal perturbation theory guarantees the completeness of the basis~(\ref{descendants-def}). But at some particular values of $\alpha$, completeness breaks down due to the operator resonance phenomenon. In the sinh\-/Gordon model these identities eliminate mixed\-/derivative descendant operators and, in a sense, generalize the equation of motion~(\ref{shG-eqmotion}). Here we study the resonance identities in the sinh\-/Gordon model in the framework of the form factor approach. This makes it possible to identify some operators by comparison with the results of perturbation theory.

Recalling the identification of the operators expressed by the reflection relations, for the exponential fields we have~\cite{Fateev:1997nn}
\eq{
\e^{\alpha\varphi(x)}=R_\alpha\e^{(Q-\alpha)\varphi(x)}=R_{-\alpha}\e^{(-Q-\alpha)\varphi(x)}.
\label{exp-reflections}
}
Here $Q=b+b^{-1}$ and $R_\alpha$ is an analytic function of $\alpha$. The function $R_\alpha$ is called reflection coefficient and is known exactly. It is related to the vacuum expectation values $G_\alpha=\langle\e^{\alpha\varphi}\rangle$ of the exponential operators:
$$
R_\alpha=G_\alpha/G_{Q-\alpha}.
$$
The  explicit form of $G_\alpha$ can be found in~\cite{Lukyanov:1996jj}, but we will not need it in what follows. The two reflections (\ref{exp-reflections}) lead to the coincidence of an infinite chain of local operators:
\eq{
\Aligned{
G_{(2k+1)Q-\alpha}\e^{\alpha\varphi(x)}
&=G_\alpha\e^{((2k+1)Q-\alpha)\varphi(x)},
\\
G_{2kQ+\alpha}\e^{\alpha\varphi(x)}
&=G_\alpha\e^{(2kQ+\alpha)\varphi(x)},
}\label{cVa-reflections}
}
where $k$ is any integer. The reflection relations (\ref{exp-reflections}) for the exponential operators generate the corresponding relations between descendant operators, though they look more complicated.

Turning our attention to the resonance identities, since the sinh\-/Gordon model can be considered as a perturbation of the Liouville theory, let us first consider the resonance identities in the latter. The Liouville field theory is defined by the action
\eq{
S_{\rm L}[\varphi]=\int d^2x\,\left({(\d_\nu\varphi)^2\over16\pi}-\mu\e^{b\varphi}\right).
\label{Liouv-action}
}
The conformal symmetry of the Liouville theory is described by two Virasoro algebras, which act in the space of local operators, with generators $L_k$, $\bar L_k$ and central charge $c=1+6Q^2$ in the standard notation. The exponential operators play the roles of primary fields. For special values of the parameter $\alpha$,
\eq{
\alpha=\alpha_{mn}\equiv{1-m\over2}b^{-1}+{1-n\over2}b,
\qquad
m,n\in\Z_{>0},
\label{alphamn-def}
}
the corresponding representations of both Virasoro algebras are degenerate. One can define two homogeneous operators $\cD_{mn}=L_{-1}^{mn}+\cdots$ and $\bar\cD_{mn}=\bar L_{-1}^{mn}+\cdots$, which generate the level $(mn,0)$ and $(0,mn)$ null\-/vectors in the Verma modules. In the Fock space they annihilate the corresponding exponents:
$$
\cD_{mn}\e^{\alpha_{mn}\varphi}=\bar\cD_{mn}\e^{\alpha_{mn}\varphi}=0.
$$
Evidently, $\cD_{mn}\bar\cD_{mn}\e^{\alpha\varphi}=0$ at the point $\alpha=\alpha_{mn}$, but if we slightly move away from this point, the operator $\cD_{mn}\bar\cD_{mn}\e^{\alpha\varphi}$ will be nonzero. Consider the $\alpha$\=/derivative of the exponential field
\eq{
\left(\varphi\,\e^{\alpha\varphi}\right)(x)={d\e^{\alpha\varphi(x)}\over d\alpha}.
\label{cV'-def}
}
Then it can be shown that the operator $\cD_{mn}\bar\cD_{mn}\left(\varphi\,\e^{\alpha_{mn}\varphi}\right)$ is a nonzero primary field. Al.~Zamolodchikov~\cite{Zamolodchikov:2003yb} found it to be proportional to the operator $\e^{\alpha_{m,-n}\varphi}$,
\eq{
\cD_{mn}\bar\cD_{mn}\left(\varphi\,\e^{\alpha_{mn}\varphi}\right)(x)
=2B_{mn}\e^{\alpha_{m,-n}\varphi(x)},
\label{resid-Liouville}
}
and calculated the proportionality coefficients~$B_{mn}$ (see Appendix~\ref{reference-appendix} for their explicit form). He called these identities the higher Liouville equations of motion, since the particular equation with $m=n=1$ is nothing but the usual equation of motion $\d\bd\varphi=2\pi\mu b\e^{b\varphi}$.

In \cite{Lashkevich:2011ne} the higher equations of motion (\ref{resid-Liouville}) were interpreted as resonance identities in the theory of a free massless boson perturbed by the field $\e^{b\varphi}$. The operator on the l.h.s.\ turns out to be the residue of an operator finitely defined in the free field theory, but possessing a pole at $\alpha=\alpha_{mn}$ in the Liouville theory. If we add an extra perturbation $\e^{-b\varphi}$, a similar consideration leads to a generalization of the identity (\ref{resid-Liouville}) to the sinh\-/Gordon model.

The form of the resonance identities in the sinh\-/Gordon theory depends on the parity of the product $mn$. In the present paper, on the basis of the form factor approach, we argue that they have the form
\eq{
\cD_{mn}\bar\cD_{mn}\left(\varphi\,\e^{\alpha_{mn}\varphi}\right)
=\Cases{2B_{mn}\left(\e^{\alpha_{m,-n}\varphi}-\e^{-\alpha_{m,-n}\varphi}\right),&\text{if $m,n\in2\Z+1$,}\\
2B_{mn}\e^{\alpha_{m,-n}\varphi}&\text{otherwise.}}
\label{resid-shG}
}
This is the main result of the present paper. In the case $m=n=1$ the identity is the equation of motion $\d\bd\varphi=4\pi\mu b\sh b\varphi$.

The equation (\ref{resid-shG}) is consistent with the result of~\cite{Lashkevich:2011ne}, where a conjectural form for odd $mn$ was derived from the general properties of the model, including the resonance condition and analytic properties of the exact vacuum expectation values of the exponential operators:
\eq{
\cD_{mn}\bar\cD_{mn}\left(\varphi\,\e^{\alpha_{mn}\varphi}\right)
=2B_{mn}\e^{\alpha_{m,-n}\varphi}
-\sum_{\substack{s\in2\Z\\0\le s<\min(m,n)}}2\cD^{(s)}_{mn}\e^{\alpha_{m,n+2m-2s}\varphi}.
\label{resid-oddmn}
}
Here $\cD^{(s)}_{mn}$ is an operator in the Heisenberg algebra of the field $\varphi(x)$ that creates an $s(m+n-s)$ level spinless descendant. The operator $\cD^{(0)}_{mn}$ is a $c$\=/number.

Our present result means that, in fact, all $\cD^{(s)}_{mn}=0$ for $s>0$ and
\eq{
\cD^{(0)}_{mn}=B_{mn}G_{m,-n}/G_{m,n+2m},
\label{C0-B}
}
where $G_{mn}=G_{\alpha_{mn}}$. Indeed, the operator $\e^{\alpha_{m,n+2m}\varphi}=G_{m,n+2m}G_{m,-n}^{-1}\e^{\alpha_{m,-n}\varphi}$ according to the reflection relations~(\ref{cVa-reflections}).

In~\cite{Lashkevich:2011ne} it was explained that this is the case if $m=1$ or $n=1$. But why can it be so for $m,n\ne1$? We think that the reason is that the operators $\cD^{(s)}_{mn}\e^{\alpha_{m,n+2m-2s}\varphi}$ must be in resonance with the operator $\e^{\alpha_{m,n+2m}\varphi}$, and are actually proportional to it. Hence, their individual contributions cannot be separated and we can assign the whole result to~$\e^{\alpha_{m,n+2m}\varphi}$. Besides, it was argued in~\cite{Lashkevich:2011ne} that the vacuum expectation value $\langle\cD_{mn}\bar\cD_{mn}\left(\varphi\,\e^{\alpha_{mn}\varphi}\right)\rangle$ should be exactly zero, if $m,n$ are odd. This fixes the coefficient~$\cD^{(0)}_{mn}$.

The case of even $mn$ has already been studied in the form factor approach in~\cite{Lashkevich:2013mca,Lashkevich:2013yja}, but the method used there cannot be applied to odd values of~$mn$. Here we propose a new representation for the form factors of the operator $\cD_{mn}\bar\cD_{mn}\left(\varphi\,\e^{\alpha_{mn}\varphi}\right)$, which makes it possible to describe both cases.

Note that the reflection relations and the $\varphi\to-\varphi$ symmetry lead to an infinite set of higher equations of motion for the descendants of every operator $\varphi\,\e^{\alpha_{mn}\varphi}$. That is, identities exist for every operator $\cD_{m'n'}\bar\cD_{m'n'}(\varphi\,\e^{\alpha_{mn}\varphi})$, where $m'=\pm m+2k>0$, $n'=\pm n+2k>0$ ($k$ integer). Explicit expressions in the general case are omitted, since they are rather cumbersome and give nothing to understanding.

The paper is organized as follows. In Sect.~\ref{ff-sec} we review the free field realization of form factors proposed in~\cite{Feigin:2008hs}. In Sect.~\ref{screening-sec} we discuss the properties of the screening operators described in~\cite{Lashkevich:2013mca,Lashkevich:2013yja} and find some important relations that involve two types of screening operators. In Sect.~\ref{identities-sec} we use these identities to obtain the identities between local operators in the sinh\-/Gordon theory. Some discussion of results and perspectives are given in Sect.~\ref{conclusion-sec}.

\section{Free field realization for form factors}
\label{ff-sec}

In principle, the dynamics of a relativistic quantum field theory on the plane is completely determined by its mass spectrum and its $S$ matrix. In the case under consideration the spectrum consists of a unique neutral boson. The relation between the mass $m$ and the parameter $\mu$ is exactly known~\cite{Zamolodchikov:1995xk}. The integrability of the sinh\-/Gordon model results in the fact that the scattering is factorized and completely described by the two\-/particle $S$ matrix $S(\theta_1-\theta_2)$, where $\theta_i$ is a rapidity of $i$th particle defined by the relations $p_i^0=m\ch\theta_i$, $p_i^1=m\sh\theta_i$. Explicitly, the $S$ matrix of the sinh\-/Gordon model is a scalar function given by
\eq{
S(\theta)={\th{1\over2}(\theta-\i\pi r)\over\th{1\over 2}(\theta+\i\pi r)},
\qquad
r={1\over1+b^2}.
\label{S-sinh}
}

We will be interested in the form factors of local operators. Let $\cO(x)$ be a local operator, $|\theta_1,\ldots,\theta_N\rangle$ be an eigenstate (defined as an in\-/state) with $N$ particles with the rapidities $\theta_1<\cdots<\theta_N$. Then the matrix elements
$$
\langle\theta_{k+1},\ldots,\theta_N|\cO(0)|\theta_1,\ldots,\theta_k\rangle
=F_\cO(\theta_1,\ldots,\theta_k,\theta_{k+1}+\i\pi,\ldots,\theta_N+\i\pi)
$$
define analytic functions $F_\cO(\theta_1,\ldots,\theta_N)$. These functions are called form factors of the local operator $\cO(x)$, and the operator $\cO(x)$ is uniquely determined by the full set of its form factors.

In the sinh\-/Gordon model the form factors of local operators has the form
\eq{
F_\cO(\theta_1,\ldots,\theta_N)
=\rho^NJ_\cO(\e^{\theta_1},\ldots,\e^{\theta_N})\prod^N_{i<j}R(\theta_i-\theta_j),
\label{FJ-rel}
}
where $\rho$ is a constant and $R(\theta)$ is an $\cO$\=/independent function. Their explicit form are given in Appendix~\ref{reference-appendix}. The functions $J_\cO(x_1,\ldots,x_N)$ are rational symmetric functions. We will often use the shorthand notation $X=(x_1,\ldots,x_N)$ for the variables and $J_\cO(X)=J_\cO(x_1,\ldots,x_N)$ for these functions.

Consider the exponential operators~$\e^{\alpha\varphi}$. It will be convenient to rewrite them in the form
\eq{
\e^{\alpha\varphi(x)}=G_\alpha V_a(x),
\qquad
a={1\over2}-{\alpha\over Q}.
\label{Va-def}
}
The $J$\=/functions corresponding to the operator $V_a(x)$ are given by~\cite{Lukyanov:1997bp}
\eq{
J_a(X)\equiv J_{V_a}(X)=\sum_{X=X_-\sqcup X_+}\e^{\i\pi(\#X_--\#X_+)a}\prod_{\substack{x\in X_-\\y\in X_+}}f\left(x\over y\right),
\label{Ja-def}
}
where
\eq{
f(x)={(x+q)(x-q^{-1})\over x^2-1},
\qquad
q=\e^{-\i\pi r}.
\label{f-def}
}
Here the sum is taken over all partitions of the set (more precisely, multiset) $X$ into two disjoint subsets $X_-$ and~$X_+$. It was shown in~\cite{Feigin:2008hs} that this function can be represented as a matrix element of special currents constructed in terms of auxiliary free bosons. Namely, consider the Heisenberg algebra generated by the elements $\d_a$, $\hat a$, $d^\pm_k$ ($k\in\Z\setminus\{0\}$) with the commutation relations
\eq{
[\d_a,\hat a]=1,
\qquad
[d^\pm_k,d^\mp_l]=kA^\pm_k\delta_{k+l,0},
\label{dpm-commut}
}
where
\eq{
A^\pm_k=(\pm)^k(q^{k/2}-q^{-k/2})(q^{k/2}-(-)^kq^{-k/2}).
\label{Apm-def}
}
All other commutators vanish. Define the vacuum vectors
\eq{
\Aligned{
&{}_a\langle1|\hat a={}_a\langle1|a,
\qquad
&&\hat a|1\rangle_a=a|1\rangle_a,
\\
&{}_a\langle1|d^\pm_{-k}=0,
\qquad
&&d^\pm_k|1\rangle_a=0
\qquad
(k>0).
}\label{vac-def}
}
The action of the Heisenberg algebra on the vacuum vectors ${}_a\langle1|$ and $|1\rangle_a$ generates the Fock spaces, which will be denoted $\cF_a=\bigoplus^\infty_{L=0}(\cF_a)_L$ and $\bcF_a=\bigoplus^\infty_{L=0}(\bcF_a)_L$ respectively. The gradation of the Fock spaces is defined naturally: $\deg d^\pm_k=k$ for $\cF_a$ and $\deg d^\pm_k=-k$ for $\bcF_a$.

The vacuum vectors define the normal ordering symbol~$\lcolon\cdots\rcolon$. We assume it to put all elements $d^\pm_k$ with $k>0$ to the right of those with $k<0$. On the other hand, it will be convenient to assume that it does not affect the order of the zero mode operators $\hat a$ and~$\d_a$, so that $\lcolon AB\rcolon$ may not coincide with $\lcolon BA\rcolon$, if the operators contain the zero mode operators.

Define the vertex operators
\eq{
\lambda_\pm(x)=\exp\sum_{k\ne0}{d^\pm_kz^{-k}\over k}.
\label{lambdapm-def}
}
Their operator products look like
\eq{
\Aligned{
\lambda_\pm(z_1)\lambda_\pm(z_2)
&=\lcolon\lambda_\pm(z_1)\lambda_\pm(z_2)\rcolon,
\\
\lambda_+(z_1)\lambda_-(z_2)
&=\lambda_-(z_2)\lambda_+(z_1)=f\left(z_2\over z_1\right)\lcolon\lambda_+(z_1)\lambda_-(z_2)\rcolon
\quad(z_2\ne\pm z_1).
}\label{lambda-norm}
}
The currents
\eq{
t(z)=\e^{\i\pi\hat a}\lambda_-(z)+\e^{-\i\pi\hat a}\lambda_+(z),
\qquad
s(z)=\lcolon\lambda_-(z)\lambda_+(-z)\rcolon.
\label{ts-def}
}
generate the algebra ${\it SVir}_{q,-q}$ described in detail in~\cite{Lashkevich:2013yja}.

Now we are ready to rewrite the r.h.s.\ of~(\ref{Ja-def}) in terms of the Heisenberg algebra. Namely,
\eq{
J_a(X)={}_a\langle1|t(X)|1\rangle_a,
\qquad
t(X)=t(x_1)t(x_2)\cdots t(x_N).
\label{Ja-ff}
}
This expression admits a simple generalization. Consider a commutative algebra $\cA$ generated by the elements $c_{-1},c_{-2},\ldots$. This algebra admits a natural gradation $\cA=\bigoplus^\infty_{L=0}\cA_L$ by assuming $\deg c_{-k}=k$. We will need two representations of the algebra $\cA$ in terms of the free bosons $d^\pm_k$:
\eq{
\pi(c_{-k})={d^-_k-d^+_k\over A^+_k},
\qquad
\bar\pi(c_{-k})={d^-_{-k}-d^+_{-k}\over A^+_k}.
\label{pibpi-def}
}
Define the vectors
\eq{
{}_a\langle h|={}_a\langle1|\pi(h),
\qquad
|h\rangle_a=\bar\pi(h)|1\rangle_a
\qquad
\forall h\in\cA.
\label{hstate-def}
}
These vectors form what we call $\cA$-subspaces in the spaces $\cF_a$, $\bcF_a$:
\eq{
\cF^\cA_a=\bigl\{{}_a\langle h|\>\big|\>h\in\cA\bigr\}\subset\cF_a,
\qquad
\bcF^\cA_a=\bigl\{|h\rangle_a\>\big|\>h\in\cA\bigr\}\subset\bcF_a.
\label{FA-def}
}
The grading of $\cA$ and those of the Fock spaces $\cF_a$ and $\bcF_a$ are consistent.

A pair of elements $h,h'\in\cA$ define a set of functions
\eq{
J^{h\bh'}_a(X)={}_a\langle h|t(X)|h'\rangle_a.
\label{Jhh'a-def}
}
By substituting these functions into~(\ref{FJ-rel}) one defines a set of functions $F^{h\bar h'}_a(\theta_1,\ldots,\theta_N)$, which satisfy the form factor axioms and, hence, define an operator~$V^{h\bh'}_a(x)$. In~\cite{Feigin:2008hs} it was argued that for generic~$a$, if $h\in\cA_L$, $h'\in\cA_{\bar L}$, the operator $V^{h\bar h'}_a(x)$ is a linear combination of the level $(L-k,\bar L-k)$ descendants of the operator $V_a(x)$ with $0\le k\le\min(L,\bar L)$ with a nonzero highest level component.

The functions $J^{h\bh'}_a(X)$ are explicitly calculated by using eqs.~(\ref{lambda-norm}), (\ref{ts-def}) together with the commutation relations
\subeq{\label{pibpi-props}
\Align{{}
[\pi(c_{-k}),\lambda_\pm(z)]
&=(\mp)^{k+1}z^k\lambda_\pm(z),
&\pi(c_{-k})|1\rangle_a
&=0,
\label{pilambda-commut}
\\
[\bar\pi(c_{-k}),\lambda_\pm(z)]
&=-(\pm)^{k+1}z^{-k}\lambda_\pm(z),
&{}_a\langle1|\bar\pi(c_{-k})
&=0,
\label{bpilambda-commut}
\\
[\pi(c_{-k}),\bar\pi(c_{-l})]
&=-(1+(-1)^k)k(A^+_k)^{-1}\delta_{kl}.
\label{pibpi-commut}
}}
Note that equation (\ref{pibpi-commut}) reflects the fact that in the massive theory the two chiralities are not independent in contrast to the conformal field theory. This simple equation implicitly underlies all our results that concern descendants with two chiralities, including the resonance identities.

In what follows we shall need the following properties of the functions~$J^{h\bh'}_a(X)$. First of all, the following quasiperiodicity property immediately follows from~(\ref{ts-def}):
\eq{
J^{hh'}_{a+1}(X)=(-1)^NJ^{hh'}_a(X).
\label{Jhh'a-quasiperiodicity}
}
Second, the function $J_a(X)$ is even in~$a$:
\eq{
J_{-a}(X)=J_a(X).
\label{Ja-reflection}
}
In the case of exponential fields $h=h'=1$ these two equations express the reflection properties~(\ref{cVa-reflections}), which can be rewritten in terms of the operator $V_a(x)$ as
\eq{
V_a(x)=V_{-a}(x)=V_{a+2}(x),
\label{Va-equations}
}
and the fact that the particle of the sinh\-/Gordon model is odd in the field~$\varphi$.

The reflection property of descendant operators is not so explicit. It was proven in~\cite{Feigin:2008hs} that there exists an analytic family $r_a$ of linear maps on $\cA$ such that
\eq{
J^{h\bh'}_a(X)=J^{r_a(h)\,\overline{r_{-a}(h')}}_{-a}(X).
\label{Jhh'a-reflection}
}
The explicit form of the reflection map $r_a$ is actually known up to the third level.

\section{Screening operators}
\label{screening-sec}

The representation (\ref{Jhh'a-def}) is valid and nondegenerate for generic values of the parameter~$a$. The nondegeneracy means that there is a one\-/to\-/one correspondence between the space of descendants over $V_a$ and the elements of the algebra $\cA\otimes\cA$, which consists of linear combination of pairs $h\bar h'=h\otimes h'$. Surely, there are special values, at which the representation is degenerate, i.~e.\ there exists an element $g\in\cA\otimes\cA$ such that $J^g_a(X)=0$. It means that in the vicinity of these special points we need to take into account derivatives ${d^k\over da^k}J^g_a(X)$ to enumerate all operators.

It was shown in~\cite{Lashkevich:2013yja} that degeneration takes place at the following values of~$a$:
\eq{
a_{mn}={rm\over2}+{(1-r)n\over2},
\qquad m,n\in\Z.
\label{amn-def}
}
These values correspond to the values $\alpha=\alpha_{mn}$, where resonances take place. From the algebraic point of view these values have the special property of being those at which the screening operators emerge. Below we show that in the form factor approach, the resonance identities arise directly due to the screening operators.

Now let us describe the algebra of screening operators introduced in~\cite{Lashkevich:2013yja}. First, define the currents
\subeq{
\label{currents-def}
\Align{
S(z)
&=\delta\lcolon\exp\sum_{k\ne0}{d^-_k-d^+_k\over k(q^{k/2}-q^{-k/2})}z^{-k}\rcolon,
\label{S-def}
\\
\tS(z)
&=\tdelta\lcolon\exp\sum_{k\ne0}{d^-_k-d^+_k\over k(\tq^{k/2}-\tq^{-k/2})}z^{-k}\rcolon,
\label{tS-def}
\\
\sigma(z)
&=\delta\lcolon\exp\sum_{k\ne0}{q^{k/2}d^-_k-q^{-k/2}d^+_k\over k(q^{k/2}-q^{-k/2})}z^{-k}\rcolon,
\label{sigma-def}
\\
\tsigma(z)
&=\tdelta\lcolon\exp\sum_{k\ne0}{\tq^{k/2}d^-_k-\tq^{-k/2}d^+_k\over k(\tq^{k/2}-\tq^{-k/2})}z^{-k}\rcolon.
\label{tsigma-def}
\\
\epsilon(z)
&=\delta\tdelta\exp\sum_{k\in2\Z+1}{2(d^-_k-d^+_k)\over k(q^k-q^{-k})}z^{-k},
\label{epsilon-def}
}}
where
\eq{
\delta=\e^{(1-r)\d_a},
\qquad
\tdelta=\e^{r\d_a},
\qquad
\tq=\e^{-\i\pi(1-r)}=\e^{-\i\pi}q^{-1}.
\label{delta-def}
}
The currents (\ref{currents-def}) together with $t(z)$ defined in~(\ref{ts-def}) generate an algebra, which will be described below. The currents $S(z)$ and $\tS(z)$ are called screening currents.

Note that there is a kind of symmetry between the currents $S(z)$, $\sigma(z)$ and $\tS(z)$, $\tsigma(z)$. The currents with tilde are obtained from those without by the substitution $r\to1-r$. To save space we will avoid duplicating relations that can be obtained one from another by this substitution.

It will be convenient to use the Laurent components of the currents. For any current $O(z)$ we will define the components $O_k$:
\eq{
O(z)=\sum_{k\in\Z}O_kz^{-k},
\qquad
O_k=\oint{dz\over2\pi\i}\,z^{k-1}O(z).
\label{Sk-def}
}
As usual, we always assume that in the product $A_kB_{k'}$ the contour of the first operator envelops all the poles of the operator product $A(z)B(z')$, while the contour of the second one leaves them outside.

The components of the currents (\ref{currents-def}) act on the Fock modules $\bcF_a$ as follows
\eq{
\Aligned{
S_k,\sigma_k:
&(\bcF_a)_L\to(\bcF_{a-(1-r)})_{L-k},
\\
\tS_k,\tsigma_k:
&(\bcF_a)_L\to(\bcF_{a-r})_{L-k},
\\
\epsilon_k:
&(\bcF_a)_L\to(\bcF_{a-1})_{L-k}.
}\label{FF-action}
}
Note that the operators $S_k$, $\tS_k$ and $\epsilon_k$ map $\cA$-subspaces into $\cA$-subspaces, so that they can be used to construct $J$-functions for local operators.

The modes $S_k$ satisfy the commutation relations
\eq{
S_kS_l=-S_{l+2}S_{k-2}.
\label{SS-commut}
}
Their commutation relations with the current $t(z)$ look like
\eq{
[S_k,t(z)]=2Bz^k\sigma(z)\cos\left(\pi\hat a-{\pi r\over2}(k-1)\right),
\label{St-commut}
}
where $B=q^{1/2}+q^{-1/2}$. To complete the picture let us give the relation between $S_k$ and~$\sigma(z)$:
\eq{
S_k\sigma(z)=z^2\sigma(z)S_{k-2}.
\label{Ssigma-commut}
}
The same relations with the substitution $r\to1-r$ (and, hence, $q\to\tq$, $B\to\tB=\tq^{1/2}+\tq^{-1/2}$) are valid for $\tS_k$,~$\tsigma(z)$.

To establish the relations that contain both types of screening currents consider the operator product
\eq{
S(z')\tS(z)={z^{\prime2}\over z^2}\tS(z)S(z')={z^{\prime2}\over z^2+z^{\prime2}}\lcolon S(z')\tS(z)\rcolon.
\label{StS-prod}
}
The residues of the poles are proportional to the following normally ordered operators:
\eq{
\lcolon S(\pm\i z)\tS(z)\rcolon=\epsilon(\tq^{\mp1/2}z),
\label{SS-epsilon-rel}
}
These two identities can be rewritten in terms of the Laurent components as follows:
\Gather{
S_k\tS_l-\tS_{l-2}S_{k+2}
={\i^l\over2}\left(q^{(k+l)/2}+(-1)^lq^{-(k+l)/2}\right)\epsilon_{k+l},
\label{StS-commut}
\\
[S_k,\epsilon_l]=[\tS_k,\epsilon_l]=0.
\label{S-epsilon-commut}
}
We see that the screening modes generally do not commute, but their `commutator' can be expressed in terms of the operators that generate conserved charges. Due to the operator products
\eq{
\epsilon(z')t(z)=t(z)\epsilon(z')={z'+z\over z'-z}\lcolon\epsilon(z')t(z)\rcolon
\label{epsilon-t-prod}
}
the operator $\epsilon(z)$ only produces a simple factor in any matrix element between $\cA$\=/states:
\eq{
{}_{a-1}\langle h|\epsilon(z)t(X)|h'\rangle_a
={}_a\langle h|t(X)|h'\rangle_a\prod^N_{i=1}{z+x_i\over z-x_i},
\qquad
h,h'\in\cA.
\label{epsilon-matrel}
}
Note that this does not mean that the modes $\epsilon_k$ commute with $t(z)$. Due to the pole in~(\ref{epsilon-t-prod}) we have
\eq{
[\epsilon_k,t(z)]=2z^k\lcolon\epsilon(z)t(z)\rcolon.
\label{epsilon-t-commut}
}
This identity will be used later. The commutation relation between the $\tS_k$ and $\sigma(z)$ will be produced later, after some more notation has been introduced.

Now specialize to the Fock spaces $\cF_{mn}=\cF_{a_{mn}}$, $\bcF_{mn}=\bcF_{a_{mn}}$. Define two operators
\Align{
\Sigma|_{\bcF_{mn}}
&=\oint{dz\over2\pi\i}\,z^{m-n}S(z)=S_{m-n+1},
\label{Sigma-def}
\\*
W|_{\bcF_{mn}}
&=\oint{dz_1\over2\pi\i}\oint{dz_2\over2\pi\i}\,
z_1^{m-n+2}z_2^{m-n}S(z_1)S(z_2)F^n(z_2/z_1)
=\sum^\infty_{k=1}F^n_kS_{m-n+3-k}S_{m-n+1+k},
\label{W-def}
}
where $F^n(z)$ is a formal series of the form
\eq{
F^n(z)=\sum^\infty_{k=1}F^n_kz^k=\sum^\infty_{k=1}(-)^{k-1}{q^{k/2}-(-)^nq^{-k/2}\over q^{k/2}+(-)^nq^{-k/2}}\,z^k.
\label{Fn-def}
}
The operators satisfy the relations
\eq{
\Sigma^2=0,
\qquad
[\Sigma,W]=0.
\label{SigmaW-commut}
}
They are used as building blocks of the screening operators defined as
\eq{
Q^{(s)}=\Cases{W^{s/2},&\text{if $s\in2\Z$;}\\
\Sigma W^{(s-1)/2},&\text{if $s\in2\Z+1$.}
}\label{Q-gen:def}
}
In particular, $\Sigma=Q^{(1)}$ and $W=Q^{(2)}$ are the two simplest screening operators. The screening operators act on the Fock spaces as
\eq{
Q^{(s)}:(\bcF_{mn})_l\to(\bcF_{m,n-2s})_{l-s(m-n+s)}.
\label{Q-gen-action}
}
Besides, they map the $\cA$\=/subspaces into $\cA$\=/subspaces. The main property of the screening operators is that they commute with the current~$t(z)$:
\eq{
[Q^{(s)},t(z)]|_{\bcF_{mn}}=0,\quad\text{if $s-n\in2\Z$.}
\label{Qt-commut}
}
The definition of the operators $\tSigma$, $\tW$, $\tQ^{(s)}$ is evident. Note however that $\tQ^{(s)}$ acts on the spaces $\bcF_{mn}$ as follows
\eq{
\tQ^{(s)}:(\bcF_{mn})_l\to(\bcF_{m-2s,n})_{l-s(n-m+s)}.
\tag{\ref{Q-gen-action}a}
}

Looking at (\ref{StS-commut}) we see that the only cases when $S_k$ and $\tS_l$ commute are those for which $l$ is odd and $k+l=0$. Hence
\eq{
[\Sigma,\tSigma]|_{\bcF_{mn}}=0,
\quad
\text{if $m,n\in2\Z+1$.}
\label{Sigma-tSigma-commut}
}
It can be also shown (see Appendix~\ref{Sigma-tW-commut-appendix}) that
\eq{
[\Sigma,\tW]|_{\bcF_{mn}}
=\Cases{0,&m\in2\Z,\\\i^{n-m-1}\epsilon_0\tSigma,&m\in2\Z+1.}
\label{Sigma-tW-commut}
}
It is assumed that $n\in2\Z+1$. The commutator $[\tSigma,W]$ looks analogously.

Eq.~(\ref{St-commut}) can be interpreted as saying that the operators $S_k$ transform the current $t(z)$ into the current $\sigma(z)$, which, in contrast to~$t(z)$, has no physical meaning. It turns out that the operators $\tS_k$ are able to transform the current $\sigma(z)$ back to $t(z)$, but for some particular values of~$a$. Namely,
\eq{
\tSigma\sigma(z)-z^{-2}\sigma(z)\tSigma|_{\bcF_{mn}}
=(-1)^{m+1\over2}B^{-1}z^{n-m-1}\lcolon\epsilon(z)t(z)\rcolon|_{\bcF_{mn}},
\quad\text{if $m\in2\Z+1$.}
\label{tSigma-sigma-commut}
}
It is easily proven from the operator product
\eq{
\sigma(z)\tS(z')={z^2\over z^{\prime2}}\tS(z')\sigma(z)
={z^2\over(z-\tq^{1/2}z')(z-\tq^{-1/2}z')}\lcolon\sigma(z)\tS(z')\rcolon
\label{sigma-tS-prod}
}
and the identity
\eq{
\lcolon\sigma(z)\tS(\tq^{\pm1/2}z)\rcolon=\lcolon\epsilon(z)\lambda_\pm(z)\rcolon.
\label{sigma-tS-lambda-rel}
}
We will use the property (\ref{tSigma-sigma-commut}) while deriving operator identities in the next section.

\section{Operator identities}
\label{identities-sec}

Now we apply the above algebra to the derivation of the resonance identities~(\ref{resid-shG}). Their l.h.s.\ must be a $m\times n$ level spinless descendant of the operator $\varphi\e^{\alpha_{mn}\varphi}$. We will search the corresponding $J$\=/functions in the form ${}_a\langle N_{mn}|t(X)|h'\rangle_a$ in the vicinity of the point $a=a_{mn}$, where $N_{mn}\in\cA_{mn}$ is such an element that the vector ${}_{mn}\langle N_{mn}|$ is a singular vector, while $h'$ is a more or less generic element in~$\cA_{mn}$. As we discussed in~\cite{Lashkevich:2013mca} the singular vector represents the descendant $\cD_{mn}\e^{\alpha_{mn}\varphi}$, which is the null vector. This singular vector is explicitly expressed in terms of the Macdonald polynomials of the elements~$c_k$, but it is more convenient to write it in terms of the screening operators. There are two representations of the same singular vector:
$$
\Aligned{
{}_{mn}\langle N_{mn}|&\sim{}_{m,-n}\langle1|Q^{(n)},
\\
{}_{mn}\langle N_{mn}|&\sim{}_{-m,n}\langle1|\tQ^{(m)}.
}
$$
The existence of these two forms reflects the $b\leftrightarrow b^{-1}$ symmetry of the model~\cite{Koubek:1993ke} and is analogous to the symmetry of screening operators in the conformal field theory~\cite{Dotsenko:1984nm}. In \cite{Lashkevich:2013mca} we used the first representative for ${}_a\langle N_{mn}|$, which made it possible to find the representative for even values of $mn$ only. Below we use the second representative. The arbitrary vector $|h'\rangle_a$ will be chosen in the form $Q^{(m)}|1\rangle_{a+2m(1-r)}$.

More precisely, denote by $Q^{(s)}_{mn}$, $\tQ^{(s)}_{mn}$ the operators $Q^{(s)}$, $\tQ^{(s)}$ defined according to rules described in the previous section so as if they act on the space $\bcF_{mn}$, but we allow them to act on any space~$\bcF_a$. Let $m,n>0$. Define the functions
\eq{
I^{(mn)}_a(X)={}_{a-rm}\langle1|\tQ^{(m)}_{mn}t(X)Q^{(m)}_{m,m+2n}|1\rangle_{a+(1-r)m}.
\label{Imn-def}
}
It is important that the vectors ${}_{a-rm}\langle1|\tQ^{(m)}_{mn}$ and $Q^{(m)}_{m,m+2n}|1\rangle_{a+(1-r)m}$ lie in the spaces $\cF^\cA_a$ and $\bcF^\cA_a$ correspondingly on the same level $mn$. It means that, being substituted into (\ref{FJ-rel}) in the place of~$J_\cO$, the functions $I^{(mn)}_a(X)$ produce the form factors of a level $mn$ descendant of the operator~$V_a(x)$, which will be denoted~$U^{(mn)}_a(x)$.

Now we want to specialize this function to the value $a=a_{mn}$. If either $m$ or $n$ is even, the matrix element $I_{mn}=I^{(mn)}_{a_{mn}}$ is given by
\eq{
I_{mn}(X)
=I_{mn}J_{m,-n}(X),
\label{Imn-even}
}
where
\eq{
I_{mn}\equiv I_{mn}(\varnothing)=(-1)^{m(n-m+1)\over2}{\left\lfloor m\over2\right\rfloor!\left\lfloor n+m\over2\right\rfloor!\over\left\lfloor n\over2\right\rfloor!}.
\label{Kmn-even}
}
In operator form these identities read
\eq{
U_{mn}(x)=I_{mn}V_{m,-n}(x),
\quad\text{if $mn\in2\Z$,}
\label{Umn-even}
}
where $U_{mn}(x)=U^{(mn)}_{a_{mn}}(x)$. Up to a factor the r.h.s.\ coincides with that of (\ref{resid-shG}) for even~$mn$. The l.h.s.\ is an $mn$ level descendant, which is consistent with $\cD_{mn}\bar\cD_{mn}\left(\varphi\,\e^{\alpha_{mn}\varphi}\right)$. We identify
\eq{
\cD_{mn}\bar\cD_{mn}\left(\varphi\,\e^{\alpha_{mn}\varphi}\right)(x)={2B_{mn}G_{m,-n}\over I_{mn}}U_{mn}(x),
\quad\text{if $mn\in2\Z$.}
\label{DDV'-U-even}
}

As we already said, identities equivalent to (\ref{Umn-even}), (\ref{DDV'-U-even}) were derived in~\cite{Lashkevich:2013mca,Lashkevich:2013yja}, but there the function
$$
{}_{m,-n}\langle1|Q^{(n)}t(X)Q^{(m)}|1\rangle_{m,n+2m}
$$
was used, instead of $I_{mn}(X)$. In fact, it is proportional to $I_{mn}(X)$ since the two representatives of the vector ${}_{mn}\langle v|$ are proportional. For even values of~$mn$ the proportionality coefficient is readily given by
\eq{
{}_{m,-n}\langle1|Q^{(n)}
={}_{-m,n}\langle1|\tQ^{(m)}{\kappa_{mn}\over I_{mn}}.
\label{sing-id}
}
where the coefficients
\eq{
\kappa_{mn}={}_{m,-n}\langle1|Q^{(m+n)}|1\rangle_{m,n+2m}=\Cases{s!\prod^s_{i=1}F^n_{2i-1},&\text{if $m+n=2s\in2\Z$,}\\
(-1)^ss!\prod^s_{i=1}F^n_{2i},&\text{if $m+n=2s+1\in2\Z+1$,}}
\label{kappa-explicit}
}
where conjectured in~\cite{Lashkevich:2013yja} and can be checked by a direct calculation (see Appendix~\ref{kappa-appendix}).

If both $m$ and $n$ are odd, from (\ref{Sigma-tSigma-commut}) and (\ref{Sigma-tW-commut}) we immediately conclude that $I_{mn}(X)=0$. From the general philosophy of what we should do in the degenerate case, we expect that the necessary level $mn$ descendant can be expressed in the form of a derivative with respect to $a$,
$$
I'_{mn}(X)=\left.{d\over da}I^{(mn)}_a(X)\right|_{a=a_{mn}}.
$$
Indeed, this derivative turns out to have the form
\eq{
I'_{mn}(X)=\pi K_{mn}\left(J_{m,-n}(X)-J_{m,n+2m}(X)\right),
\label{Imn-odd}
}
where
\eq{
K_{mn}=-(-1)^{n-1\over2}\,{{m-1\over2}!{m+n-2\over2}!\over{n-1\over2}!}.
\label{Kmn-odd}
}
In operator notation eq.~(\ref{Imn-odd}) reads
\eq{
U'_{mn}(x)=\pi K_{mn}\left(V_{m,-n}(x)-V_{m,n+2m}(x)\right),
\qquad m,n\in2\Z+1.
\label{Umn-odd}
}
Here $U'_{mn}(x)$ corresponds to the functions $I'_{mn}(X)$ as the $J$\=/functions. By comparing it with (\ref{resid-oddmn}) we conclude that, if we identify
\eq{
\cD_{mn}\bar\cD_{mn}\left(\varphi\,\e^{\alpha_{mn}\varphi}\right)(x)={2B_{mn}G_{m,-n}\over\pi K_{mn}}U'_{mn}(x),
\quad\text{if $mn\in2\Z$,}
\label{DDV'-U-odd}
}
the coefficients $\cD^{(s)}_{mn}$ in the r.h.s.\ of (\ref{resid-oddmn}) vanish for all $s>0$, while $\cD^{(0)}_{mn}$ is given by~(\ref{C0-B}).

Now let us produce the detailed derivation of the identities (\ref{Imn-even}) and~(\ref{Imn-odd}).

\subsection{Even level identities}
\label{identities-even-subsec}

Consider first the case of even $m\times n$. To calculate $I_{mn}(X)$ notice that in this case $\tQ^{(m)}$ in (\ref{Imn-def}) commutes with $t(X)$. Hence,
$$
I_{mn}(X)={}_{-m,n}\langle1|t(X)\tQ^{(m)}Q^{(m)}|1\rangle_{m,n+2m}
={}_{-m,n}\langle1|t(X)|1\rangle_{-m,n}\times{}_{-m,n}\langle1|\tQ^{(m)}Q^{(m)}|1\rangle_{m,n+2m}.
$$
The first factor on the r.h.s.\ is nothing but $J_{-m,n}(X)=J_{m,-n}(X)$, which is the `functional' part of the r.h.s.\ of eq.~(\ref{Imn-even}). The last factor is constant and can be calculated explicitly. Here we give some more general expressions, since we shall need them later. In Appendix~\ref{WW-vme-appendix} we prove that
\Multline{
J^{(s)}_{mn}\equiv{}_{m-2s,n}\langle1|\tQ^{(s)}Q^{(s)}|1\rangle_{m,n+2s}
=(-1)^{(n-m+1)s\over2}\,{{s\over2}!\left\lfloor n-m+2s\over2\right\rfloor!\over\left\lfloor n-m+s\over2\right\rfloor!},
\\
\text{if $s\in2\Z$ and $n-m+s\ge0$.}
\label{WW-vme}
}
For even values of $m$ we have $I_{mn}=J^{(m)}_{mn}$, which immediately proves~(\ref{Kmn-even}). In the case of odd $m$ and even $n$ we have
\Multline{
{}_{-m,n}\langle1|\tQ^{(m)}Q^{(m)}|1\rangle_{m,n+2m}
={}_{-m,n}\langle1|\tQ^{(m-1)}\tSigma\Sigma Q^{(m-1)}|1\rangle_{m,n+2m}
\\
={}_{-m,n}\langle1|\tQ^{(m-1)}\Sigma\tSigma Q^{(m-1)}|1\rangle_{m,n+2m}
+(-1)^{n-m+1\over2}{}_{-m,n}\langle1|\tQ^{(m-1)}\epsilon_0Q^{(m-1)}|1\rangle_{m,n+2m}.
\notag
}
The first term on the second line vanishes, since, due to~(\ref{Sigma-tW-commut}),
$$
\tSigma Q^{(m-1)}|1\rangle_{m,n+2m}=Q^{(m-1)}\tSigma|1\rangle_{m,n+2m}
=Q^{(m-1)}\tS_{m+n+1}|1\rangle_{m,n+2m}=0.
$$
In the second term the operator $\epsilon_0$ commutes with $Q^{(m-1)}$ and then only shifts the zero mode of the ket\-/vacuum. Hence,
\Multline{
{}_{-m,n}\langle1|\tQ^{(m)}Q^{(m)}|1\rangle_{m,n+2m}
=(-1)^{n-m+1\over2}{}_{-m,n}\langle1|\tQ^{(m-1)}Q^{(m-1)}|1\rangle_{m-2,n+2m-2}\\
=(-1)^{n-m+1\over2}J^{(m-1)}_{m-2,n}=I_{mn},
\notag
}
q.e.d.

\subsection{Odd level identities: the \texorpdfstring{$m=1$}{m=1} case}
\label{identities-odd-1n-subsec}

Our next aim is to consider the case of odd values of both $m,n$ and to prove the identity~(\ref{Imn-odd}). Since the general proof is rather cumbersome, let us start with the simplest case $m=1$ to better explain the idea.

Evidently, by taking into account the fact that in the matrix element (\ref{Imn-def}) only the $t$\=/currents are $a$\=/dependent, and using the product rule, we obtain in the $m=1$ case
$$
I'_{1n}(X)=\sum^N_{i=1}{}_{-1,n}\langle1|\tQ^{(1)}t'(x_i)t(\hat X_i)Q^{(1)}|1\rangle_{1,n+2}
=\sum^N_{i=1}{}_{-1,n}\langle1|\tSigma t'(x_i)t(\hat X_i)\Sigma|1\rangle_{1,n+2}.
$$
Here $\hat X_i=X\setminus\{x_i\}$. Since $t'(x_i)$ commutes with every $t(x_j)$, we put it to the left. The screening operators are explicitly $\tSigma|_{\bcF_{1n}}=\tS_n$, $\Sigma|_{\bcF_{1,n+2}}=S_{-n}$.

First pull $\Sigma$ to the left. It commutes with $t(\hat X_i)$ and $\tSigma$, but catches on~$t'(x_i)$. Due to (\ref{St-commut}) we have
$$
[t'(x),\Sigma]|_{\bcF_{1,n+2}}=2\pi(-1)^{n+1\over2}Bz^{-n}\sigma(z)\quad(n\in2\Z+1).
$$
Then
$$
I'_{1n}(X)
=2\pi(-1)^{n+1\over2}B\sum^N_{i=1}x_i^{-n}\times{}_{-1,n}\langle1|\tSigma\sigma(x_i)t(\hat X_i)|1\rangle_{1,n+2}.
$$
Now we want to push the operator $\tSigma$ to the right. Again, it commutes with $t(\hat X_i)$, but catches on~$\sigma(x_i)$. By applying eq.~(\ref{tSigma-sigma-commut}) for $m=1$ we obtain
\eq{
I'_{1n}(X)
=2\pi(-1)^{n-1\over2}\sum^N_{i=1}{}_{-1,n}\langle1|\lcolon\epsilon(x_i)t(x_i)\rcolon t(\hat X_i)|1\rangle_{1,n+2}.
\label{I-epsilonnormprod}
}
Eq.~(\ref{epsilon-t-prod}) makes it possible to represent $\lcolon\epsilon(x_i)t(x_i)\rcolon$ as ${1\over2}[\epsilon_0,t(z)]$. For matrix elements between $\cA$\=/states it means:
\Multline{
\sum^N_{i=1}{}_{a-1}\langle h|\lcolon\epsilon(x_i)t(x_i)\rcolon t(\hat X_i)|h'\rangle_a
={}_{a-1}\langle h|{1\over2}[\epsilon_0,t(X)]|h'\rangle_a
\\
={1\over2}\left({}_a\langle h|t(X)|h'\rangle_a-{}_{a-1}\langle h|t(X)|h'\rangle_{a-1}\right)
={1-(-1)^N\over2}\>{}_a\langle h|t(X)|h'\rangle_a.
\label{epsilon_0-matrel}
}
Substituting this in~(\ref{I-epsilonnormprod}) we finally obtain
\eq{
I'_{1n}(X)=(-1)^{n-1\over2}\pi(J_{1,n+2}(X)-J_{-1,n}(X)).
\label{IJ-rel}
}
This proves (\ref{Imn-odd}) in the case $m=1$. In the particular case $n=1$ we have got the fourth (and the simplest) proof, in the framework of the form factor approach, of the identity equivalent to the equation of  motion. Another three proofs can be found in~\cite{Babujian:2002fi,Feigin:2008hs,Lashkevich:2013yja}.

\subsection{Odd level identities: the general case}
\label{idenitites-odd-subsec}

Consider now the general matrix element
$$
I'_{mn}(X)
=\sum^N_{i=1}{}_{-m,n}\langle1|\tW^{m-1\over2}\tSigma t'(x_i)t(\hat X_i)\Sigma W^{m-1\over2}|1\rangle_{m,n+2m}.
$$
Pushing the operator $\Sigma$ to the left and taking into account the fact that it commutes with $\tQ^{(m)}$, we obtain
\eq{
I'_{mn}(X)
=2\pi\i^{n+1}B
\sum^N_{i=1}x_i^{m-n-1}\times{}_{-m,n}\langle1|\tW^{m-1\over2}\tSigma\sigma(x_i)t(\hat X_i)W^{m-1\over2}|1\rangle_{m,n+2m}.
\label{Iprime-tWtSigma-intermed}
}
Now push the operator $\tSigma$ to the right. Using the commutation relation (\ref{Sigma-tW-commut}) we get
\Align{
I'_{mn}(X)
&=\pi\i^{n-m}
{}_{-m,n}\langle1|\tW^{m-1\over2}[\epsilon_0,t(X)]W^{m-1\over2}|1\rangle_{m,n+2m}
\notag
\\
&\quad
-2\pi\i^m\left(m-1\over2\right)B\sum^N_{i=1}x_i^{n-m-3}
\times{}_{-m,n}\langle1|\tW^{m-1\over2}\sigma(x_i)t(\hat X_i)\Sigma W^{m-3\over2}|1\rangle_{m-2,n+2m-2}.
\notag
}
Pushing again $\Sigma$ to the left in the second term we obtain it in the form similar to~(\ref{Iprime-tWtSigma-intermed}):
\Align{
I'_{mn}(X)
&=(1-(-1)^N)\pi\i^{n-m}\times
{}_{-m+2,n+2}\langle1|\tW^{m-1\over2}t(X)W^{m-1\over2}|1\rangle_{m,n+2m}
\notag
\\
&\quad
+2\pi\i^{n+1}\left(m-1\over2\right)^2B\sum^N_{i=1}x_i^{m-n-5}
\times{}_{-m+2,n+2}\langle1|\tW^{m-3\over2}\tSigma\sigma(x_i)t(\hat X_i)W^{m-3\over2}|1\rangle_{m-2,n+2m-2}
\notag
\\
&=(1-(-1)^N)\pi\i^{n-m}J^{(m-1)}_{mn}(X)+\left(m-1\over2\right)^2I'_{m-2,n+2}(X),
\notag
}
where we used the notation
\eq{
J^{(s)}_{mn}(X)={}_{m-2s,n}\langle1|\tW^{s/2}t(X)W^{s/2}|1\rangle_{m,n+2s}.
\label{Js-def}
}
Evidently, $J^{(s)}_{mn}(\varnothing)=J^{(s)}_{mn}$. Taking (\ref{IJ-rel}) as initial condition for the iteration procedure we get
\eq{
I'_{mn}(X)
=\sum_{\substack{s\in2\Z\\0\le s\le m-1}}(-1)^{n-s-1\over2}\pi\left({m-1\over2}!\over{s\over2}!\right)^2
\left(J^{(s)}_{m,n+2m-2s}(X)-J^{(s)}_{-m,n-2s}(X)\right).
\label{IJgen-rel}
}
The functions $J^{(s)}_{mn}(X)$ correspond to a level $s(n-m+s)$ spinless descendant operator over~$V_{mn}(x)$. But this is not the end of the story. Checking the resonance conditions we see that this operator may occur in resonance with the exponential operators $V_{m,n+2s}(x)$ and~$V_{-m,n-2m+2s}(x)$. In the form factor construction it reveals itself in the possibility to reduce the functions $J^{(s)}_{mn}(X)$ to a linear combination of $J_{m,n+2s}(X)$ and~$J_{-m,n-2m+2s}(X)$. Below we show that
\eq{
J^{(s)}_{mn}(X)=(-1)^{s/2}{{s\over2}!{n-m+2s-2\over2}!\over{n-m+s\over2}!}
\left({n-m+s\over2}J_{m,n+2s}(X)+{s\over2}J_{-m,n-2m+2s}(X)\right),
\label{Jsmn-explicit}
}
if $n\ge m-s$. With this result eq.~(\ref{IJgen-rel}) reduces to~(\ref{Imn-odd}), q.e.d.

To prove (\ref{Jsmn-explicit}) first of all pull $W^{s/2}$ in the r.h.s.\ of~(\ref{Js-def}) to the left. We have
\Align{
J^{(s)}_{mn}(X)
&={}_{m-2s,n}\langle1|\tW^{s/2}W^{s/2}|1\rangle_{m,n+2s}\times
{}_{m,n+2s}\langle1|t(X)|1\rangle_{m,n+2s}+C^{(s)}_{mn}(X)
\notag
\\
&=J^{(s)}_{mn}I_{m,n+2s}(X)+C^{(s)}_{mn}(X),
\label{Jsmn-comm-def}
}
where
\Align{
C^{(s)}_{mn}(X)
&={}_{m-2s,n}\langle1|\tW^{s/2}[t(X),W^{s/2}]|1\rangle_{m,n+2s}
\notag
\\
&=\sum^{s/2}_{j=1}{}_{m-2s,n}\langle1|\tW^{s/2}W^{j-1}[t(X),W]W^{s/2-j}|1\rangle_{m,n+2s}
\notag
\\
&=-2\i^nB\sum^{s/2}_{j=1}\sum^N_{i=1}x_i^{m-n-1-4j}
\times{}_{m-2s,n}\langle1|\tW^{s/2}W^{j-1}\Sigma\sigma(x_i)t(\hat X_i)W^{s/2-j}|1\rangle_{m,n+2s}
\notag
\\
&=-\i^n sB\sum^N_{i=1}x_i^{m-n-1}
\times{}_{m-2s,n}\langle1|\tW^{s/2}\sigma(x_i)t(\hat X_i)\Sigma W^{s/2-1}|1\rangle_{m,n+2s}.
\label{Jsmn-comm}
}
We used the fact that $\Sigma W^k$ commutes with $t(z)$ for any~$k$. Now push $\Sigma$ to the left and commute it with~$\tW^{s/2}$:
$$
C^{(s)}_{mn}(X)
=(-1)^{m+1\over2}{s^2\over2}B\sum^N_{i=1}x_i^{m-n-3}
\times{}_{m-2s,n}\langle1|\tW^{s/2-1}\tSigma\sigma(x_i)t(\hat X_i)W^{s/2-1}|1\rangle_{m,n+2s}.
$$
Pushing $\tSigma$ to the right we get
\Align{
C^{(s)}_{mn}(X)
&=\left(s\over2\right)^2(1-(-1)^N)\times{}_{m-2s+4,n+4}\langle1|\tW^{s/2-1}t(X)W^{s/2-1}|1\rangle_{m,n+2s}
\notag
\\
&\quad
+\i^nB\left(s\over2\right)^2(s-2)\sum^N_{i=1}x_i^{m-n-5}
\times{}_{m-2s+4,n+4}\langle1|\tW^{s/2}\sigma(x_i)t(\hat X_i)\Sigma W^{s/2-1}|1\rangle_{m,n+2s}
\notag
\\
&=\left(s\over2\right)^2\left((1-(-1)^N)J^{(s-2)}_{m,n+4}(X)-C^{(s-2)}_{m,n+4}(X)\right).
\notag
}
The last line follows from the last line of (\ref{Jsmn-comm}) and, together with (\ref{Jsmn-comm-def}), provides a recurrence relation for $J^{(s)}_{mn}(X)$:
\eq{
J^{(s)}_{mn}(X)
=J^{(s)}_{mn}J_{m,n+2s}(X)
-\left(s\over2\right)^2\left((-1)^NJ^{(s-2)}_{m,n+4}(X)-J^{(s-2)}_{m,n+4}J_{m,n+4}(X)\right).
\label{Jsmn-recurrence}
}
The function $J^{(s)}_{mn}$ is given explicitly by~(\ref{WW-vme}). The initial condition for the recurrence relation is provided by the evident fact that $J^{(0)}_{mn}(X)=J_{mn}(X)$. By using the fact that $J_{-m,n-2m}(X)=(-1)^NJ_{mn}(X)$ we easily check that the r.h.s.\ of~(\ref{Jsmn-explicit}) solves the recurrence relation.

\section{Conclusion}
\label{conclusion-sec}

We derived the identities (\ref{resid-shG}) between the operators defined in terms of exact bootstrap form factors that correspond to resonance identities in the perturbation theory of the sinh\-/Gordon model. We calculated form factors in the framework of the algebraic approach and used the current algebra. To do it we needed to calculate the derivatives of form factors of a descendant field in the parameter $\alpha$ (or, equivalently, $a$). The novel technical tool, which makes it possible to find such derivatives, was the use of the relation~(\ref{tSigma-sigma-commut}). It allowed us to get rid of the currents $\sigma(z)$, in combination with the relations~(\ref{Sigma-tSigma-commut}) and~(\ref{Sigma-tW-commut}), and to reduce some complicated expressions, which contained $W$ and $\tW$, to $J$-functions of exponential fields. In particular, this drastically simplifies the derivation of the equation of motion. A generalization of this construction to the sine\-/Gordon model is an open question.

A generalization of the relation~(\ref{tSigma-sigma-commut}) turns out to have some important applications. In particular, it makes it possible to study form factors of the conserved currents and their products. Another field  of application is to select the local operators consistent with the reductions. Results dealing with these subjects will be published soon.

\section*{Acknowledgments}

We are grateful to H.~Babujian, I.~Marshall, G.~Mussardo, F.~Smirnov for discussions. Study of the algebra of screening operators was performed with the support of the Russian Scientific Foundation under the grant \#~14--12--01383.

\Appendix

\section{Some reference data}
\label{reference-appendix}

Here, for reference purposes we produce some explicit expressions omitted in the main text.

The coefficients $B_{mn}$ in the higher equations of motion (\ref{resid-Liouville}) and the resonance identities~(\ref{resid-oddmn}), (\ref{resid-shG}) are given by~\cite{Zamolodchikov:2003yb}
\eq{
B_{mn}=b^{1+2n-2m}\left(\mu{\pi\Gamma(b^2)\over\Gamma(1-b^2)}\right)^n{\Gamma(n-mb^2)\over\Gamma(1-n+mb^2)}
\prod_{\substack{1-m\le k\le m-1\\1-n\le l\le n-1\\(k,l)\ne(0,0)}}(kb^{-1}+lb).
\label{Bmn}
}
The constant $\rho$ and the function $R(\theta)$, which enter the form factors according to~(\ref{FJ-rel}), are~\cite{Lukyanov:1997bp}
\eq{
\Gathered{
\rho=\left(2\cos{\pi r\over2}\right)^{-1/2}\exp\int^{\pi(1-r)}_0{dt\over2\pi}\,{t\over\sin t},
\\
R(\theta)=\exp\left(
-4\int^\infty_0{dt\over t}\,
{\sh{\pi t\over2}\sh{\pi(1-r)t\over2}\sh{\pi rt\over2}
\over\sh^2\pi t}\ch(\pi-\i\theta)t
\right).
}\label{R-def}
}

\section{Proof of~(\ref{Sigma-tW-commut})}
\label{Sigma-tW-commut-appendix}

Let us calculate $[\Sigma,\tW]$ for odd values of~$n$. The calculation of $[\tSigma,W]$ repeats it literally, and we omit it. By using the definition of $\tW$ and the commutation relation (\ref{StS-commut}) we obtain
\Multline{
[\Sigma,\tW]|_{\bcF_{mn}}
={\i^{m-n-3}\over2}\sum^\infty_{k=1}\tF^m_k\Bigl(
  \left(\tq^{-k/2}+(-1)^m\tq^{k/2}\right)\epsilon_{-k}\tS_{n-m+1+k}
\\
  -\left(\tq^{k/2}+(-1)^m\tq^{-k/2}\right)\tS_{n-m+1-k}\epsilon_k
\Bigr)
\\
={\i^{m-n-1}\over2}\sum_{k\ne0}(-1)^{k-1}\left(\tq^{k/2}+(-1)^m\tq^{-k/2}\right)\epsilon_k\tS_{n-m+1-k}
\\
={\i^{m-n-1}\over2}\sum_{k\in\Z}(-1)^{k-1}\left(\tq^{k/2}+(-1)^m\tq^{-k/2}\right)\epsilon_k\tS_{n-m+1-k}
+{\i^{m-n-1}\over2}(1-(-1)^m)\epsilon_0\tSigma.
\label{Sigma-tW-deriv}
}
We used the fact that $\epsilon_k$ commutes with~$\tS_l$. The sum in the last line reduces to the integral
\Multline{
-\oint{dz\over2\pi\i}z^{n-m}\left(\epsilon(-\tq^{-1/2}z)\tS(z))-(-1)^m\epsilon(-\tq^{1/2}z)\tS(z)\right)
\\
=-\delta^2\tdelta^2\oint{dz\over2\pi\i}z^{n-m}\left(S^+(\i z)-(-1)^mS^+(-\i z)\right)
=-\i^{n-m-1}(1+(-1)^n)\delta^2\tdelta^2S^+(z)=0,
\notag
}
where $S^+(z)=\lcolon S^{-1}(z)\rcolon$. Here we used the fact that
\eq{
\epsilon(-\tq^{\pm1/2}z)\tS(z)=\delta^2\tdelta^2S^+(\mp\i z).
\label{epsilon-tS=S+}
}
Hence, the first term in the last line of~(\ref{Sigma-tW-deriv}) vanishes, while the second one provides the r.h.s.\ of~(\ref{Sigma-tW-commut}).

\section{Calculation of the coefficients~\texorpdfstring{$\kappa_{mn}$}{k(mn)}}
\label{kappa-appendix}

Here we prove eq.~(\ref{kappa-explicit}). Let $m+n=2s$. Then $Q^{m+n}=W^s$ and we have
\Multline{
\kappa_{mn}={}_{m,m-2s}\langle1|W^s|1\rangle_{m,m+2s}
\\
=\sum_{k_1,\ldots,k_s\ge1}\prod^s_{i=1}F^m_{k_i}\times
{}_{m,m-2s}\langle1|\prod^{\substack{\curvearrowright\\s}}_{i=1}S_{2s+1-2i-k_i}
\prod^{\substack{\curvearrowleft\\s}}_{i=1}S_{-2s-1+2i+k_i}|1\rangle_{m,m+2s}.
\\
\notag
}
The terms on the r.h.s.\ can only be nonzero if, for all $i$, $k_i\le2s-1$, otherwise one or other of the corresponding screening operators that contains it kills one of the vacuum vectors. Besides, the integers $k_i$ must be odd. Indeed, if any $k_i$ is even, the mode $S_{2s+1-2i-k_i}$ being commuted to the left produces the state $\langle1|S_{2s-1-k_i}$ of the degree $2s-1-k_i$, which is an odd integer. Every term that enters this state contains $c_{-k}$ with odd $k$. The corresponding operator $\pi(c_{-k})$ commutes with every $S_l$ and, hence, kills the ket\-/vacuum~$|1\rangle$.

Lastly, the terms that contain $k_i=k_j$ vanish, since by commuting the operators we get a product of the form $S_{2s+1-2l-k_i}S_{2s+1-2l-k_j}$, which is zero. Altogether, (1)~$k_i$ are odd, (2)~$1\le k_i\le 2s-1$, and (3)~all $k_i$ are different. Therefore, nonzero terms are enumerated by transpositions $l\in\mathbb S_s$, so that $k_i=2l_i-1$. All $s!$ nonzero terms coincide and equal to
$$
\prod^s_{i=1}F^n_{2i-1}\times{}_{m,m-2s}\langle1|S_0^{2s}|1\rangle_{m,m+2s}.
$$
The matrix element here is equal to one, so that $\kappa_{mn}=s!\prod^s_{i=1}F^n_{2i-1}$, which proves the first case of~(\ref{kappa-explicit}).

Now let $m+n=2s+1$. Then
$$
\kappa_{mn}={}_{m,m-2s-1}\langle1|W^s\Sigma|1\rangle_{m,m+2s+1}={}_{m,m-2s-1}\langle1|W^sS_{-2s}|1\rangle_{m,m+2s+1}.
$$
By the same argument we reduce it to
\Multline{
\kappa_{mn}=s!\prod^s_{i=1}F^n_{2i}\times
{}_{m,m-2s-1}\langle1|S_0^sS_2^sS_{-2s}|1\rangle_{m,m+2s+1}
\\
=(-1)^ss!\prod^s_{i=1}F^n_{2i}\times
{}_{m,m-2s-1}\langle1|S_0^{2s+1}|1\rangle_{m,m+2s+1}
=(-1)^ss!\prod^s_{i=1}F^n_{2i},
\notag
}
which proves the second case.

\section{Proof of~(\ref{WW-vme})}
\label{WW-vme-appendix}

First, write down the operators $W$, $\tW$ explicitly:
\Multline{
{}_{m-2s,n}\langle1|\tW^{s/2}W^{s/2}|1\rangle_{m,n+2s}
=\sum_{\{0\le k_i,l_i\le K\}}\prod^{s/2}_{i=1}\tF^m_{K+1-k_i}F^n_{K+1-l_i}
\\
\times{}_{m-2s,n}\langle1|\prod^{\substack{\curvearrowright\\s/2}}_{i=1}\tS_{k_i+2-2i}
\prod^{\substack{\curvearrowleft\\s/2}}_{i=1}\tS_{2K-k_i-2s+2i+2}
\prod^{\substack{\curvearrowright\\s/2}}_{i=1}S_{l_i-2K+2s-2i-2}
\prod^{\substack{\curvearrowleft\\s/2}}_{i=1}S_{-l_i-2+2i}|1\rangle_{m,n+2s}.
\label{WW-explicit}
}
Here $K=n-m+2s-2$. The summation variables $k_i$, $l_i$ ($i=1,\ldots,s/2$) run over all nonnegative integers, but we may restrict the sum to even values, because of the reason explained in Appendix~\ref{kappa-appendix}. The terms with $k_i=k_j$ or $l_i=l_j$ ($i\ne j$) vanish due to the identities $\tS_k\tS_{k-2}=S_kS_{k-2}=0$. Moreover, the expression under the sum is symmetric with respect to permutations of~$k_i$ and those of~$l_i$. Hence the summation sign in~(\ref{WW-explicit}) can be substituted by
$$
\left({s\over2}!\right)^2\sum_{\substack{0\le k_1<k_2<\cdots<k_{s/2}\le K\\k_i\in2\Z}}\>
\sum_{\substack{0\le l_1<l_2<\cdots<l_{s/2}\le K\\l_i\in2\Z}}
$$
Now it is time to push $\tS_k$ to the right and $S_k$ to the left. It can be shown that after that they kill the right and left vacuum vectors respectively. Hence, only their commutators will contribute the matrix element. But the commutator of $\tS_k$ and $S_l$ is proportional to $\epsilon_{k+l}$, which commutes with everything. It kills the right vacuum if $k+l>0$, and kills the left vacuum if $k+l<0$. It means that only the terms with $k_i=l_i$ will contribute the sum, and the corresponding matrix elements are all equal to~$(-1)^{(n-m)s/2}$. Therefore,
$$
{}_{m-2s,n}\langle1|\tW^{s/2}W^{s/2}|1\rangle_{m,n+2s}
=(-1)^{(n-m)s\over2}\left({s\over2}!\right)^2\sum_{\substack{0\le k_1<k_2<\cdots<k_{s/2}\le K\\k_i\in2\Z}}
\prod^{s/2}_{i=1}\tF^m_{K+1-k_i}F^n_{K+1-k_i}.
$$
However, the product $\tF^m_{K+1-k_i}F^n_{K+1-k_i}$ is equal to $-1$ for~$k_i$ even giving the total sign factor $(-1)^{s/2}$. The cardinal number of the set of admissible values of $(k_1,\ldots,k_{s/2})$ is equal to~${\lfloor K/2\rfloor+1\choose s/2}$, which proves~(\ref{WW-vme}).

\raggedright
\bibliographystyle{mybib}
\bibliography{main}

\end{document}